\documentclass[12pt,a4paper,dvips]{article}
\usepackage{a4p}
\usepackage{cite,mcite}
\usepackage{graphicx}
\usepackage{physics}
\usepackage{rotating}
\usepackage{l3_titlePH,ifthen,Lep}

\journalname{Phys. Lett. B}

\def\ZG{\ensuremath{\Zo\gamma}}

\def\EEZG{\ensuremath{\ee\ra\ZG}}
\def\EEFFG{\ensuremath{\ee\ra\ffbar\gamma}}

\def\QQG{\ensuremath{\qqbar\gamma}}
\def\NNG{\ensuremath{\nnbar\gamma}}
\def\ZGQQG{\ensuremath{\Zo\gamma\ra\qqbar\gamma}}
\def\ZGNNG{\ensuremath{\Zo\gamma\ra\nnbar\gamma}}
\def\EEQQG{\ensuremath{\ee\ra\qqbar\gamma}}
\def\EEQQGG{\ensuremath{\ee\ra\qqbar\gamma(\gamma)}}
\def\EENNG{\ensuremath{\ee\ra\nnbar\gamma}}
\def\EENNGG{\ensuremath{\ee\ra\nnbar\gamma(\gamma)}}
\def\ZZG{\ensuremath{\Zo\Zo\gamma}}
\def\ZGG{\ensuremath{\Zo\gamma\gamma}}
\def\ZVG{\ensuremath{\Zo V\gamma}}
\def\THFCM{\ensuremath{\theta_f^\Zo}}
\def\PHFCM{\ensuremath{\phi_f^\Zo}}

\newlength\leftl
\newlength\downl
\newlength\rightl
\newlength\templ
\def\spb#1#2#3#4#5{%
  \setbox0\hbox{#1}\setbox1\hbox{\char'023}%
  \rightl=#2\wd0 \advance\rightl by-#3\wd1
  \downl=#5\ht1 \advance\downl by-#4\ht0
  \leftl=\rightl \advance\leftl by\wd1
  \ht1=\downl \dp1=-\downl
  \leavevmode
  \kern\rightl\lower\downl\box1\kern-\leftl #1}
\journalname{Phys. Lett. B}
\date{April 19, 2004}
\preprint{2004-014}
\graphicspath{{/l3/paper/example/figs/}}
\newlength{\capindent}
\newlength{\capwidth}
\newlength{\figwidth}
\newcommand{\icaption}[2][!*!,!]{\hspace*{\capindent}%
  \begin{minipage}{\capwidth}
    \ifthenelse{\equal{#1}{!*!,!}}%
      {\caption{#2}}%
      {\caption[#1]{#2}}
  \end{minipage}}
\begin{document}
\begin{titlepage}

\title{Study of the \mbox{\boldmath $\epem \ra \ZG$} Process at LEP and \\
  Limits on Triple Neutral-Gauge-Boson Couplings}
\author{The L3 Collaboration}
%
%
\begin{abstract}
\indent

 The process $\epem \ra \Zo\gamma$, where the $\Zo$ boson decays into
 hadrons or neutrinos, is studied with
 data collected with the L3 detector at LEP at centre-of-mass energies
 from $189\GeV$ up to $209\GeV$. The cross sections are measured and
 found to be in agreement with the Standard Model predictions. Limits on
 triple neutral-gauge-boson couplings, forbidden in the Standard Model at tree
 level, are derived. Limits on the energy scales at which the
 anomalous couplings could be manifest are set. They range from $0.3\TeV$
 to $2.3\TeV$ depending on the new physics effect
 under consideration.

\end{abstract}

\submitted

\end{titlepage}

%
\section{Introduction}

The process $\epem \ra \Zo \gamma$ allows to test the existence
of new physics\cite{renard}, such as anomalous couplings
between neutral gauge bosons. Effects 
coming from $\Zo\Zo\gamma$ and $\Zo\gamma\gamma$ couplings are expected to 
be very small in the Standard Model\cite{renard,barroso}, but can be enhanced 
in compositeness models\cite{renard2} or if new particles enter in higher 
order corrections. Anomalous
$\Zo\Zo\gamma$ and $\Zo\gamma\gamma$ couplings would increase
 the $\epem \ra \Zo \gamma$ cross section and produce an
 enhancement of large polar angle photons.

Assuming only Lorentz and $U(1)_{em}$ gauge invariance, the most general 
form of the $\ZVG$ vertices, with a real $\Zo$ and $\gamma$ in the final state,
is parametrized by means of
the anomalous couplings, 
$h_i^V~(i=1 \ldots 4; V=\gamma,\Zo)$ \cite{def_coupl}.
 The couplings $h_1^V$ and $h_2^V$ are CP violating whereas $h_3^V$ and $h_4^V$
 are CP conserving.
 All these couplings are zero at tree level in the
Standard Model, and only the CP conserving ones are non-vanishing
($\sim 10^{-4}$) at the one-loop level{\protect \cite{renard,barroso}}. An
alternative parametrization, which introduces the energy scales of
new physics, $\Lambda_{iV}$, is \cite{mery}:

\begin{eqnarray}
   \frac{\sqrt{\alpha}~h_i^V}{\MZ^2} \equiv \frac{1}{\Lambda_{iV}^2} &  i=1,3
                                                            \label{for:lambda13} \\
   \frac{\sqrt{\alpha}~h_i^V}{\MZ^4} \equiv \frac{1}{\Lambda_{iV}^4} &  i=2,4,
                                                            \label{for:lambda24}
\end{eqnarray}

 \noindent
 where $\alpha$ is the fine-structure constant.
 The fact that there are always two identical particles at the vertex forbids
the three bosons to be on-shell. This means the 
 $\ZZG$ and $\ZGG$ vertices may only appear if one of the bosons is off-shell.
 A treatment of these vertices
where all three bosons are off-shell is discussed in Reference \citen{offshell}.
 In this Letter, the $\epem \ra \Zo \gamma$ process is analyzed.
 The maximal experimental sensitivity is achieved
 with the analysis of the $\EEFFG$ process, with the fermion pair
 in the vicinity of the $\Zo$ resonance,
 where the signal statistics is high and backgrounds are reduced. In this scenario,
 effects from an off-shell final-state $\Zo$ boson are negligible \cite{alcaraz}.

 L3 published results \cite{l3} on the $\epem \ra \Zo \gamma$ process,
setting limits on $\Zo\Zo\gamma$ and $\Zo\gamma\gamma$ couplings 
from data obtained at lower center-of-mass energies ($161 \GeV \le \sqrt{s} \le 189
\GeV$).
Results have also been published by other experiments at LEP
\cite{Lep} and at the TEVATRON\cite{tevatron}. In this Letter we present results
for the highest energies collected at LEP.

The phase space definition for the $\epem \ra \Zo \gamma$
process requires a photon with energy greater than $20 \GeV$
and polar angle in the range $5^\circ < \theta_\gamma < 175^\circ$. Every
cross section and acceptance in this Letter is referred to this
fiducial region .

%
\section{Data and Monte Carlo Samples}

Data collected by the L3 detector \cite{L3det} at  
$\sqrt{s} = 189\GeV - 209\GeV$ with a total luminosity of about
$626 \pb$ are used to study the $\epem \ra \Zo \gamma$ process 	
 in the channels $\EEQQG$ and $\EENNG$.

The Standard Model processes giving rise to these final states are modelled
with 
 KK2f, for $\EEQQGG$, and KKMC, for
$\EENNGG$ \cite{kk2f}.
 Both programs are general purpose Monte Carlo generators for the process
$\ee \ra\ffbar + n\gamma$, containing complete $O(\alpha^{2})$  corrections
from initial- and final-state radiation, including their interference.
  Data for $\EENNGG$ at $\sqrt{s}=189 \GeV$, previously analysed and compared
 to the KORALZ Monte Carlo\cite{KoralZ}, are now re-analysed using KKMC.

 Background processes are simulated with 
EXCALIBUR \cite{EXCALIBUR} for the four-fermion final
states, PHOJET \cite{phojet} and DIAG36 \cite{diag36} for two-photon collisions
with hadrons or leptons in the final state, respectively, and
 BHWIDE \cite{bhwide} and TEEGG \cite{TEEGG} for  $\ee \ra \ee\gamma(\gamma)$ .

 All generated events are passed 
through a simulation of the L3 detector \cite{geant} and the same 
analysis procedure as used for the data. Time-dependent detector inefficiencies,
monitored during data-taking, are also taken into account.

%
%
\section{Event Selection}

\subsection{Photon Selection }

The main signature of the process $\EEZG$ is the production of a high energy
photon. A photon candidate is
identified as a shower in the barrel or endcap region of the BGO crystal
electromagnetic calorimeter, consistent with an electromagnetic shower and
 with a minimum energy of $5 \GeV$. The mass of the system recoiling against the 
photon of energy $E_{\gamma}$:
 \mbox {$m_{rec}=(s-2E_{\gamma}\sqrt{s})^{1/2}$}, is required to satisfy
$80\GeV < m_{rec} < 110\GeV$, consistent with $\Zo$-boson production. For
the $\sqrt{s}$ values considered, the cuts on the recoiling mass correspond to
photon energies between $62\GeV$ and $89\GeV$.

%
%

\subsection{Selection of  {\boldmath $\ee\ra\qqbar\gamma$}  Events}

In addition to requiring a photon candidate, $\EEQQG$ events are selected by
demanding that:

\begin{itemize}
\item the event have more than 6 charged tracks reconstructed in the fiducial
volume of the tracking chamber and more than 11 calorimetric clusters in the 
electromagnetic calorimeter.

\item the transverse energy imbalance be less than 15\% of the 
  total reconstructed energy and the longitudinal energy imbalance less 
   than 20\% of the same quantity.

\item the angle of the photon candidate 
with respect to the beam direction, $\theta_\gamma$, satisfy $|\cos\theta_{\gamma}|<0.97$. 

\end{itemize}

 In order to reject electrons produced in the central region, photon candidates
 with
\mbox{$|\cos\theta_{\gamma}|< 0.90$} are not considered if they are
 associated to a charged track in the central tracking chamber. This requirement
 eliminates a substantial contamination of background processes. For the last
 period of data taking, corresponding to data at $\sqrt{s} > 202 \GeV$, this
 rejection cut is relaxed to 
 \mbox{$|\cos\theta_{\gamma}|< 0.85$} to account for different running conditions of the detector. 
This change increases the contamination
 from background processes.

 Table~\ref{tab:1} lists the data luminosity analysed at each $\sqrt{s}$,
 the selection efficiency, the background level and the number of selected events,
 after background subtraction.  

The trigger inefficiency is estimated to be negligible due to the redundancy 
of subtriggers involved in tagging this final state. Two   backgrounds contribute 
in equal proportions: the 
$\ee \ra {\rm q \bar{q}' e}\nu$ and 
$\ee \ra \qqbar \ee$ processes, where one electron fakes a photon.

\begin{figure}
\vspace*{-3.0truecm}
\begin{center}
\includegraphics[width=9.0truecm]
{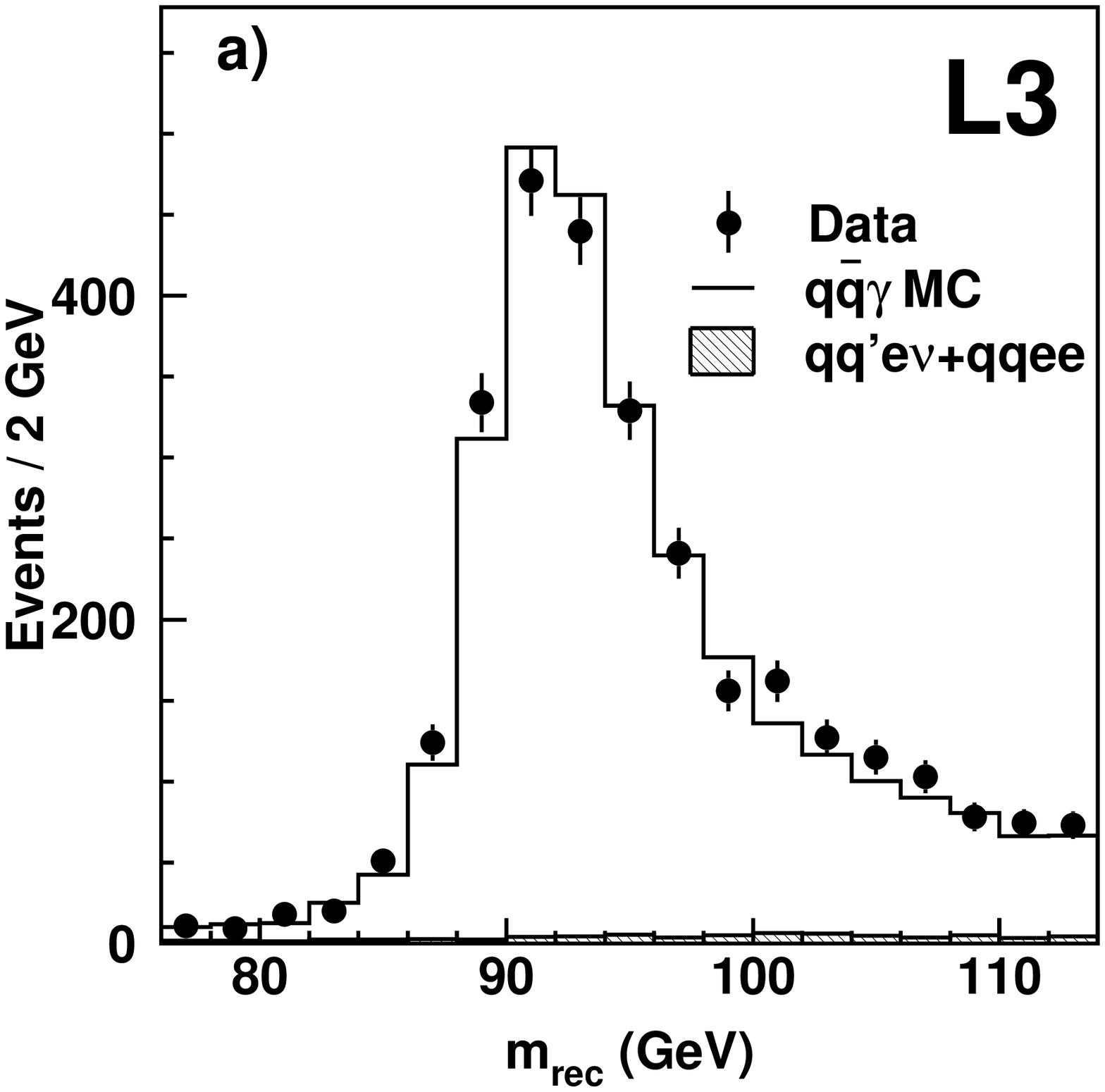} 
\begin{tabular}{lr}
\includegraphics[width=9.0truecm]
{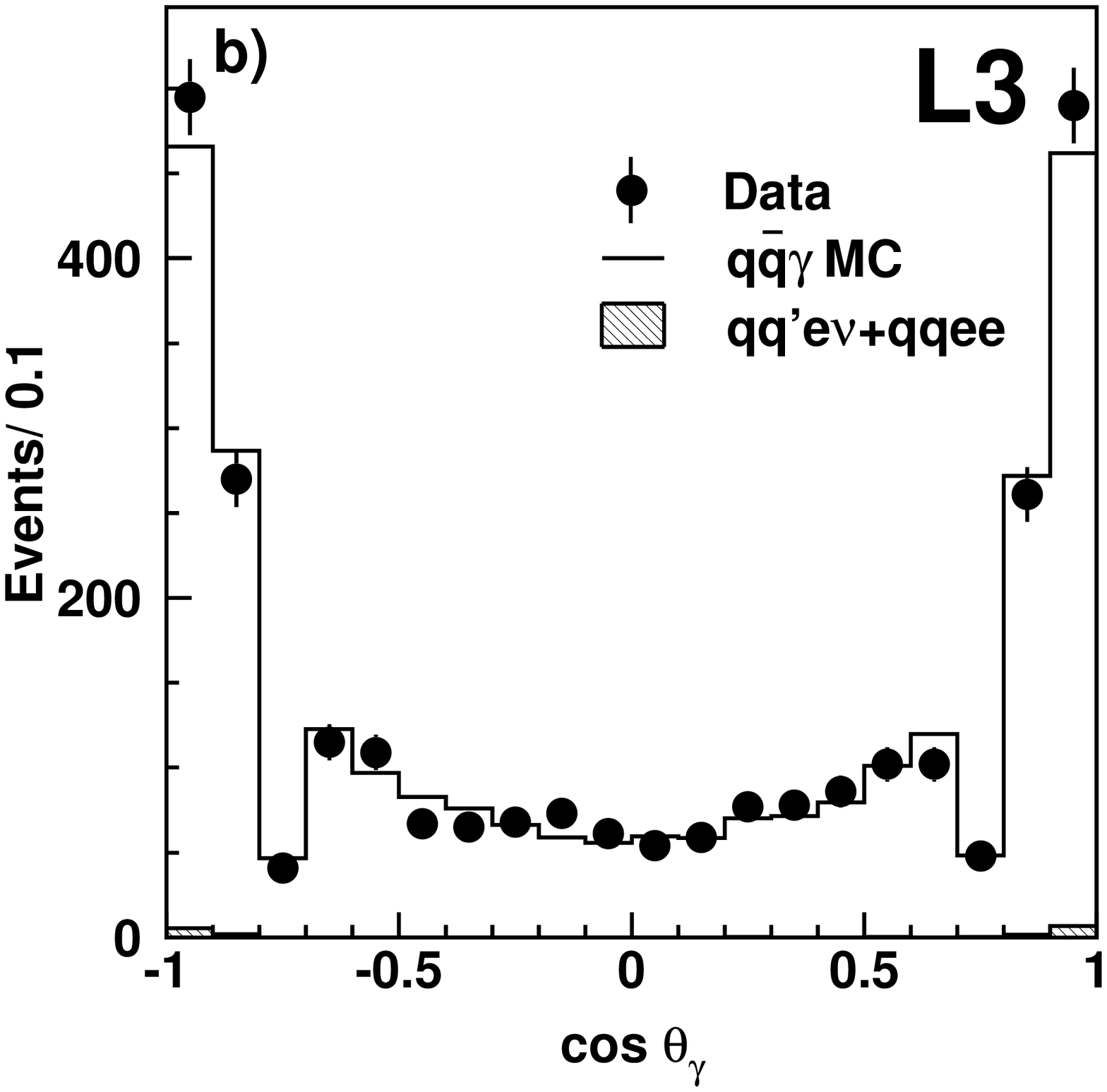} &
\includegraphics[width=9.0truecm]  
{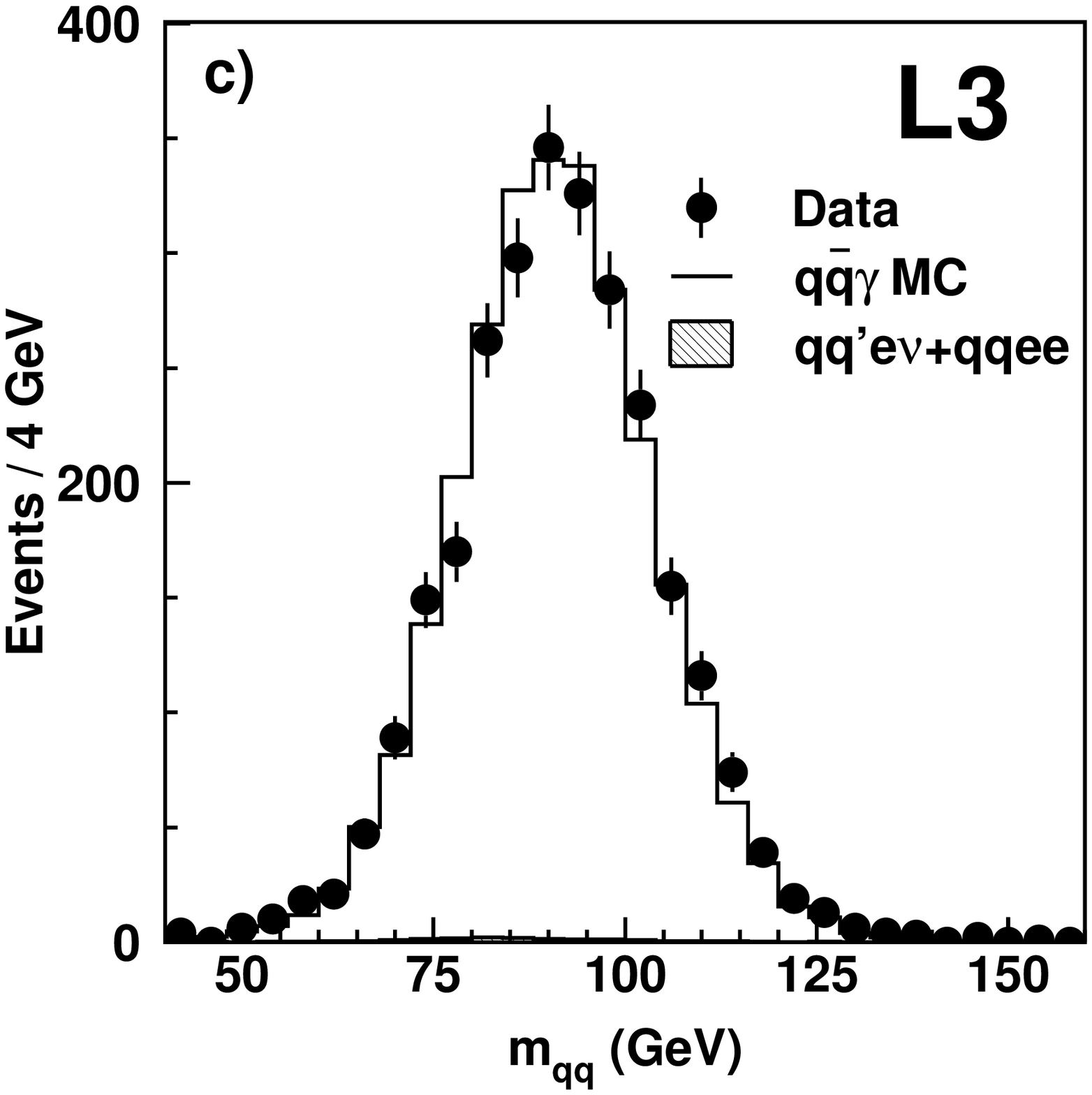}  \\
\end{tabular}
\icaption{\label{fig:1}
      Distributions of
     a) the recoil mass to the photon candidate in $\EEQQG$ events
     b) the polar angle of the photon and
     c) the invariant mass of the hadron system.
 The points are data, the open histogram is the
 Standard Model Monte Carlo prediction and the hatched one is
  the background from the
 ${\rm e}^{+}{\rm e}^{-} \ra {\rm q}\bar{\rm q'}{\rm e}\nu$ and
 ${\rm e}^{+}{\rm e}^{-} \ra {\rm q}\bar{\rm q}{\rm e}^{+}{\rm e}^{-}$
 processes.}
\end{center}
\end{figure}

 Figure \ref{fig:1} shows the distributions of $m_{rec}$, 
 $\cos{\theta_{\gamma}}$ and the invariant mass of the hadron
 system, reconstructed from jet and photon directions and $\sqrt{s}$.
 
The resolution of the L3 electromagnetic calorimeter, better than 1\%,
allows the observation of a tail in  Figure \ref{fig:1}a, for $m_{rec}$
above the nominal $\Zo$ mass, due to initial state radiation photons.

%
%
\subsection{Selection of {\boldmath $\ee\ra\nnbar\gamma$} Events}
\begin{figure}
\vspace*{-3.0truecm}
 \begin{center}
 \includegraphics[width=12.0truecm]
 {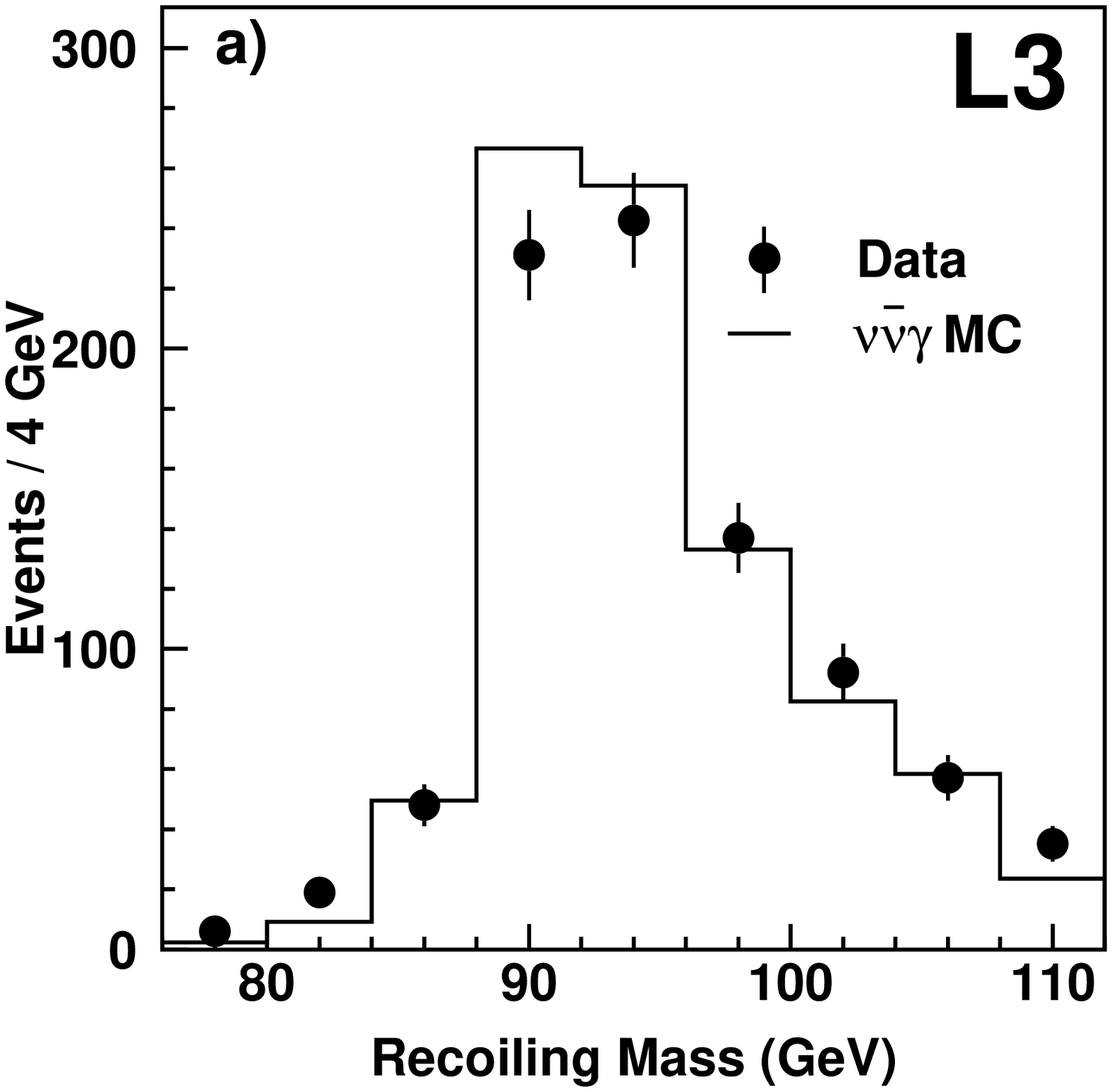}
 \includegraphics[width=12.0truecm]
 {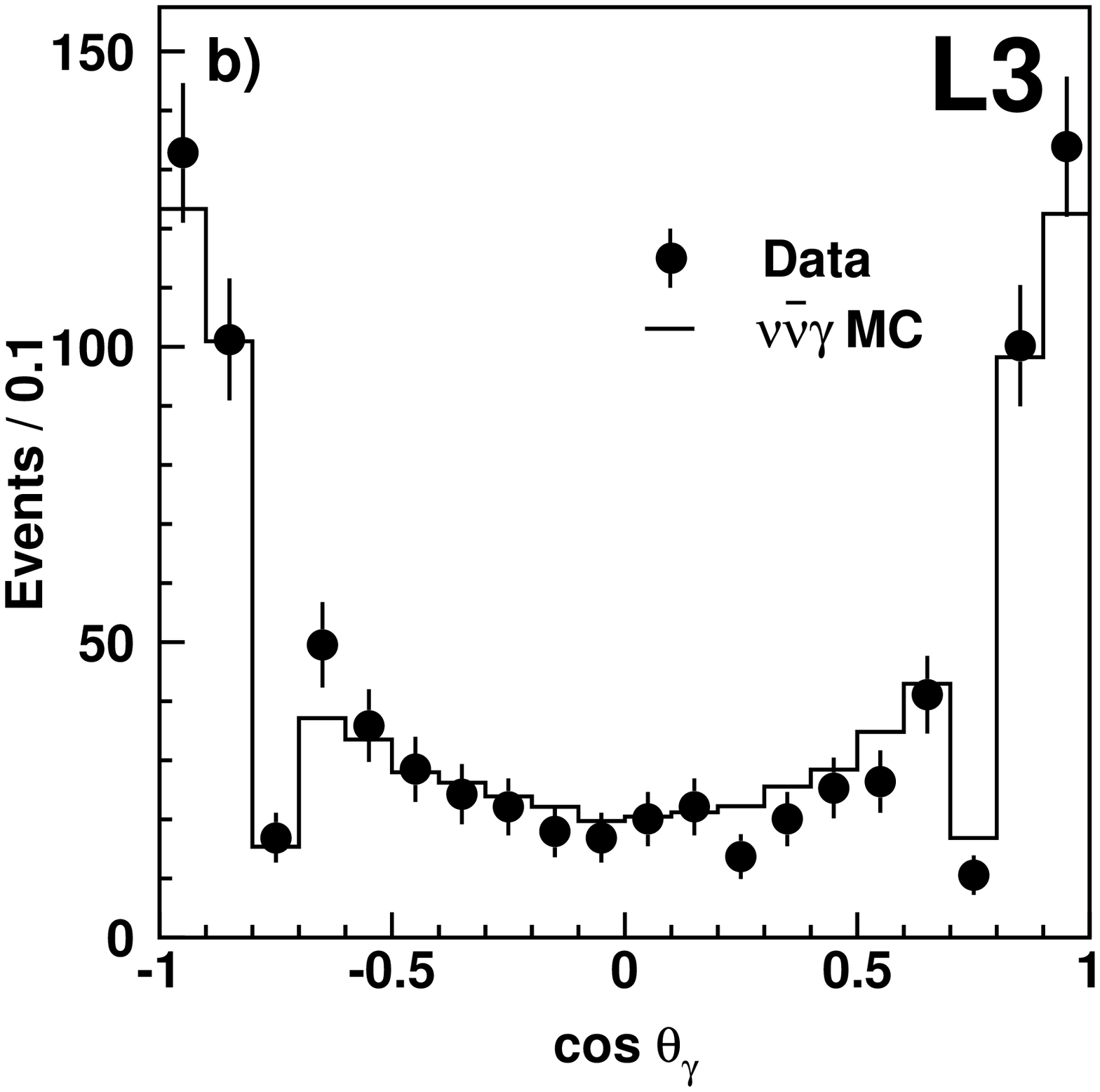}
 \icaption{\label{fig:2}
      Distributions of
     a) the recoil mass to the photon candidate in $\EENNG$ events and
     b) the photon polar angle.
     The points are data and the histogram is the Standard Model Monte Carlo
    prediction.}
 \end{center}
 \end{figure}

In addition to the presence of a photon, selected as described above,
the events from the $\EENNG$ process are selected by the following criteria:
\begin{itemize}
\item the event must have at most 5 calorimetric clusters, due to low
 energy \mbox{($< 1\GeV$)} initial state photons or noise in the calorimeter.
  The number of hits in the tracking chamber associated to a
      calorimetric cluster, including the photon candidate,
       must not exceed 40\% of the expected number of hits
      for a charged track.

\item the angle of the photon candidate
with respect to the beam direction must satisfy
    $|\cos\theta_{\gamma}|<0.96$. This range differs from that of the
   hadronic channel in order to match the angular coverage of the central
   tracking chamber used to reject electrons.

\item the total reconstructed energy, $E_{tot}$, must fulfill
  $ \sqrt{s} - E_{tot} > 0.95 \, E_{tot} $ and
  the transverse energy imbalance must be greater than $0.2 E_{tot}$.

\item To suppress cosmic ray background, there must be at least one scintillator
      time measurement within $\pm 5$ ns of the beam crossing time. The scintillator
      signals must be associated with calorimetric clusters.

\end{itemize}
The background in the selected sample is found to be negligible. 
Table \ref{tab:2} lists the data luminosity analysed at each
$\sqrt{s}$ and the selection efficiency. The selection efficiency includes
trigger efficiency, evaluated to be around 95\% by
 using two independent data samples from the $\ee\ra\ee$ and $\ee\ra\gamma\gamma$
 processes. The number of selected events is also given.
Figure \ref{fig:2} shows
the distributions of $m_{rec}$ and $\cos{\theta_{\gamma}}$.

%
\section{Cross Section Measurements}

  The measured cross sections for both the  $\EEQQG$ and $\EENNG$ processes are
 presented in Tables \ref{tab:1} and \ref{tab:2}, respectively,
 together with Standard Model predictions. 
 Good agreement is observed.
 The uncertainty on the expected cross section, $\sigma_{SM}$,
 takes into  account a 1\% theory
 uncertainty of KK2f and KKMC 
 and the finite Monte Carlo statistics generated for these studies.

 In addition to cross sections,  in Table \ref{tab:3} we present more detailed information for the
 $\EEQQG$ process on the
  number of events observed and expected, the background and selection efficiencies, in bins of
 $m_{rec}$ and $|\cos{\theta_{\gamma}}|$. Similar tables are given elsewhere
 \cite{marat} for the $\EENNG$ process.

The main sources of systematic uncertainties are summarized in Table 
\ref{tab:4}.
 The largest contribution is due to the selection procedure. A change of 3\%
in the values of the cut on $m_{rec}$ corresponds to 0.8\% and 1.5\%
uncertainties for hadronic and invisible decay modes, respectively.
  Changes in the photon energy scale give uncertainties of 0.4\% and 0.6\%
 for the  hadronic and invisible channels, respectively.
    The uncertainty from limited Monte Carlo statistics amounts to
 0.4\% for the $\EENNG$ channel and varies between 0.1\% and 0.4\% for the
 $\EEQQG$ channel.
   The accuracy on the luminosity estimation gives a 0.2\% uncertainty.
  Uncertainty in the measurement of the trigger efficiency  contributes an
 additional 0.3\% in the $\EENNG$ process.
   A variation of 10\% in the background level corresponds to a 0.3\% 
 uncertainty in the hadronic channel.

\begin{figure}
\begin{center}
\includegraphics[width=15.0truecm]
{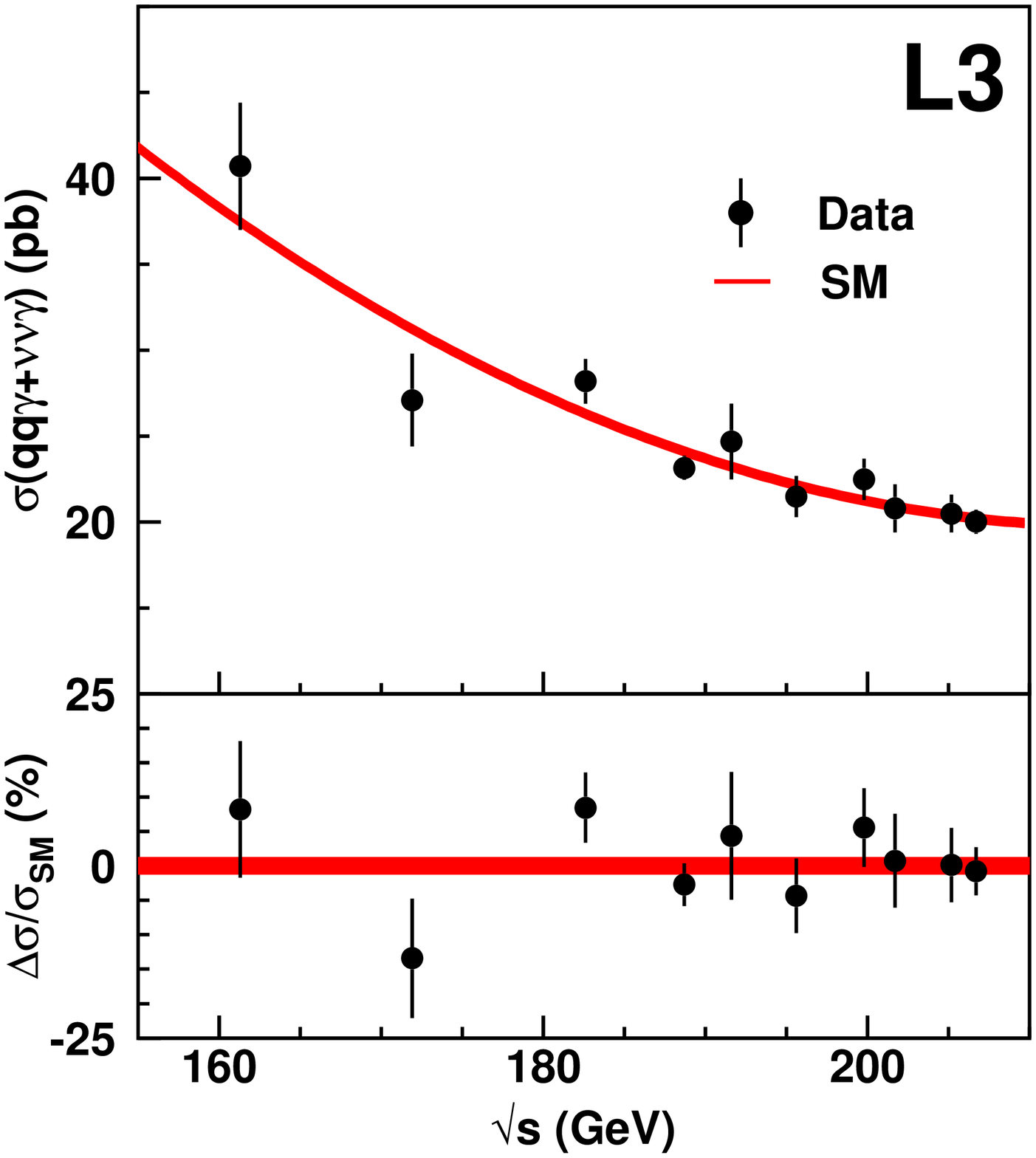}
\icaption{\label{fig:3}
          Variation of the sum of the cross-sections of the $\EEQQG$ and $\EENNG$
         processes,  $\sigma(\QQG + \NNG)$, with $\sqrt{s}$.
          Data are represented by the dots, while the solid line
          gives the variation of the Standard Model cross-section,
          $\sigma_{\rm SM}$, as calculated with the KK2f and
          KKMC {\protect \cite{kk2f}} Monte Carlo programs.
         The width of the band takes into
          account a 1\% uncertainty in each of the $\EEQQG$ and $\EENNG$ 
          theoretical cross sections. 
          The lower plot shows the relative difference.}
\end{center}
\end{figure}

The variation of the sum of the $\EEQQG$ and $\EENNG$ cross sections
with $\sqrt{s}$
 is presented in Figure \ref{fig:3}.
 Cross sections at $\sqrt{s} = 161, 172$ and $183 \GeV$, already published by
 L3 \cite{l3},  are included for completeness.
 The relative deviation from the Standard Model predictions 
  as a function of $\sqrt{s}$ is also shown. Good agreement is found.


\section{Triple Neutral-Gauge-Boson Couplings}
 Since deviations from Standard Model expectations are found neither for the
 $\EEQQG$ nor for the $\EENNG$ process,
 limits on anomalous triple-neutral-gauge boson couplings are extracted
 by using an optimal observable method\cite{opob}.

\subsection{Optimal Observable Method}

  In the presence of anomalous couplings, the cross section for the
 process $\epem \ra \Zo \gamma$ is proportional to
  $| M_{SM} + M_{AC}|^{2}$, with
  $ M_{SM}$ and $M_{AC}(h_{i}^{V})$
 the Standard Model and anomalous coupling amplitudes, respectively.

  As $M_{AC}$ depends linearly on the $h_{i}^{V}$ (i=1 \ldots 4; V=$\gamma, \Zo$)
  parameters,  the differential cross section can be written as a quadratic
  function on the anomalous couplings:
\begin{eqnarray}
\frac{d\sigma}{d\vec{\Omega}}=c_{\scriptscriptstyle 0}(\vec{\Omega})+
\sum^{4}_{i=1}\sum_{V=\gamma,\Zo} c_{\scriptscriptstyle 1,i,V}(\vec{\Omega}) h_{i}^{V}+
\sum^{4}_{i=1}\sum_{V=\gamma,\Zo}\sum^{4}_{j=1}\sum_{V'=\gamma,\Zo}c_{\scriptscriptstyle 2,ij,V,V'}(\vec{\Omega})
h_{i}^{V}h_{j}^{V'}
\end{eqnarray}
\noindent
where $\vec{\Omega}$ stands for the phase space variables defining the final
 state, $c_{\scriptscriptstyle 0}(\vec{\Omega})$ is the Standard Model cross 
 section and
 $c_{\scriptscriptstyle 1,i,V}$ and $c_{\scriptscriptstyle 2,ij,V,V'}$
 are coefficients related to the anomalous amplitudes.

   The variables defined for each coupling as:

\begin{eqnarray}\phantom{0}
{\cal O}_{1,i,V}(\vec{\Omega}) & \equiv &
 \frac{c_{\scriptscriptstyle 1,i,V}(\vec{\Omega})}
{c_{\scriptscriptstyle 0}(\vec{\Omega})}
\end{eqnarray}

\begin{eqnarray}\phantom{0}
{\cal O}_{2,ij,V,V'}(\vec{\Omega}) &
 \equiv & \frac{c_{\scriptscriptstyle 2,ij,V,V'}(\vec{\Omega})}
{c_{\scriptscriptstyle 0}(\vec{\Omega})}
\end{eqnarray}

\noindent
  are called ``optimal variables'' as they contain the full
 kinematic  information on the event and allow the determination of
 the parameters $h_{i}^{V}$ with the maximum possible statistical precision.
  If the parameters $h_{i}^{V}$ are sufficiently small, the
quadratic term can be neglected and
all the information in the multidimensional phase space 
$\vec{\Omega}$ is projected into the variable 
${\cal O}_{1}$.

   In order to extract the $h_{i}^{V}$'s, a
 binned maximum-likelihood fit of the expected distribution of the optimal variable
 ${\cal O}_{1,i,V}$ is performed to the data,
 assuming a Poisson density distribution in each bin.
 Both the shape of the optimal variable distribution, which includes
 energy and angular information, and the total number of events contribute to
  the fit.
 The  expected number of events in the presence of anomalous couplings is
 computed from a Standard Model reference sample by applying a reweighting
 technique, in which each Monte Carlo event is weighted with the following
 quantity, defined at the generator level:
\begin{eqnarray}\phantom{0}
  W(h_{i}^{V}) = \frac{| M_{SM} +  M_{AC}(h_{i}^{V})|^{2}}
                  {| M_{SM}|^{2}}
\end{eqnarray}
  The comparison between expected and observed events
  is done at the level of
 reconstructed variables so that all experimental effects, such as
 detector resolution or selection efficiencies, are automatically taken into
 account.


\subsection{Limits on Anomalous Couplings}

  Making use of the optimal observable method and
 taking into consideration the information on the total event rate for each
 process and the phase space variables defining the final state,
 limits at 95\% Confidence Level (CL) are set on the $h_{i}^{V}$ couplings.
   The reconstructed set of variables used to compute the optimal variables is
$\vec{\Omega}=(E_\gamma, \theta_\gamma, \phi_\gamma, \THFCM, \PHFCM)$, where $E_\gamma, \theta_\gamma$ 
and $\phi_\gamma$ are the energy and angles of the
photon, and $\THFCM$ and $\PHFCM$ the angles of the fermion $f$ in the $\Zo$ rest 
frame. In the $\EENNG$ channel only the three photon variables are used.

\begin{sidewaysfigure}
\begin{center}
 \begin{tabular}{lr}
\includegraphics[width=9.0truecm]{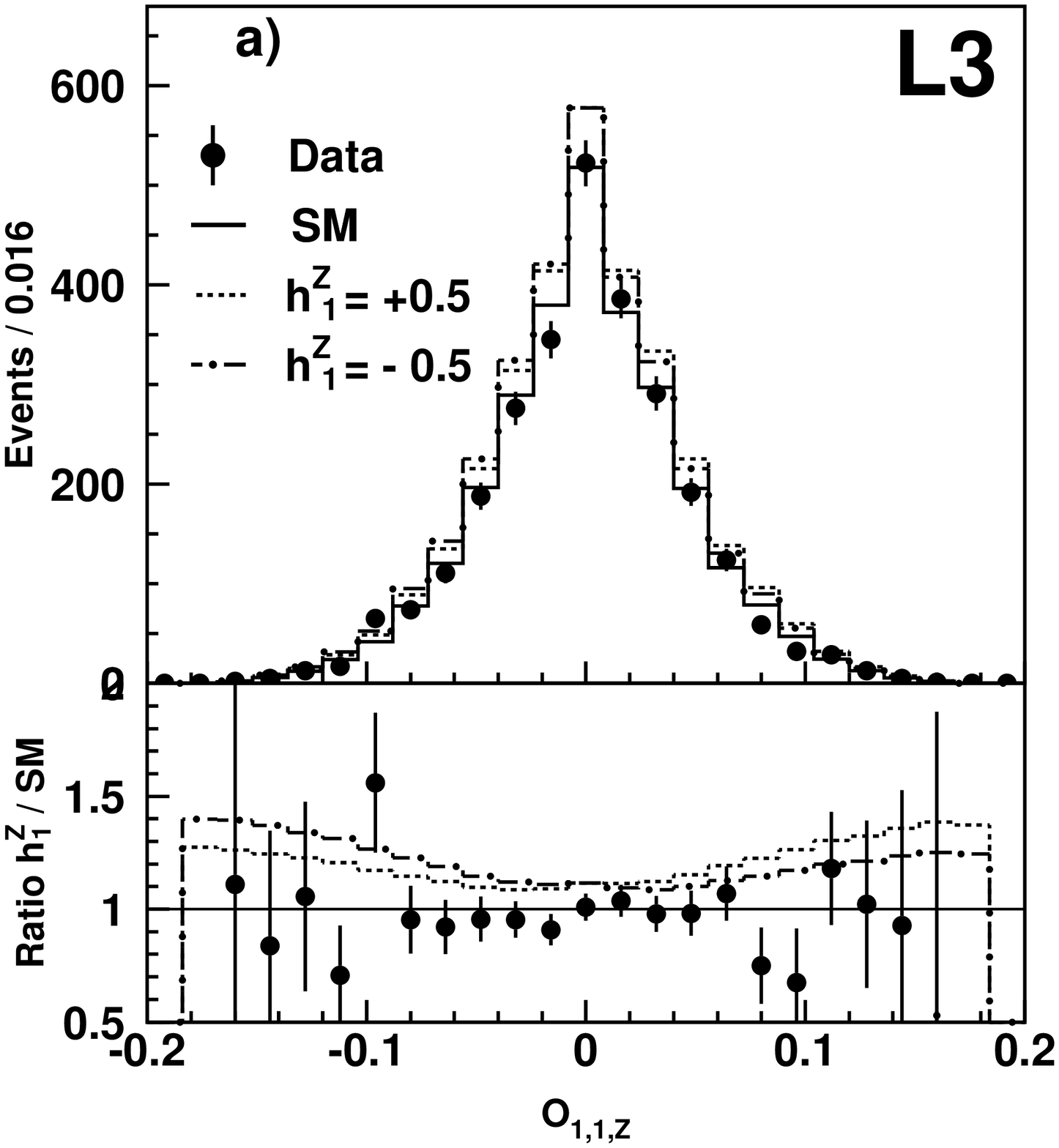} 
\includegraphics[width=9.0truecm]{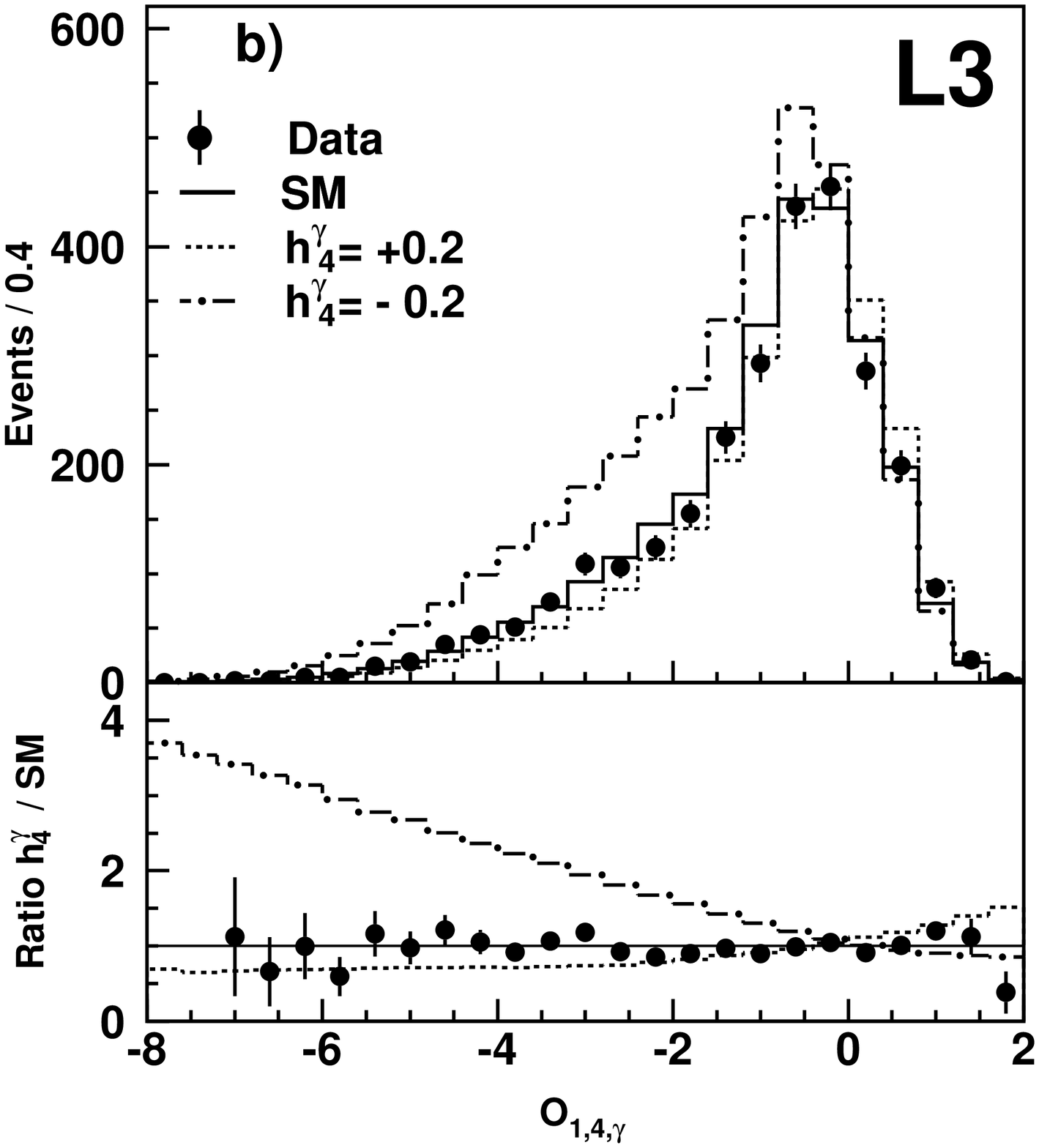} 
 \end{tabular}
\caption{\label{fig:4}
          Distribution of the optimal variables for a) the CP-violating
    coupling $h_{1}^{\Zo}$ and b) for the CP-conserving coupling $h_{4}^{\gamma}$.
          Data are shown together with the 
          expectations for the Standard Model (SM) and for values of
           anomalous couplings $h_{1}^{\Zo}=\pm 0.5$ and $h_{4}^{\gamma}=\pm 0.2$.
          The lower plots shows the ratios between the
          anomalous coupling contributions and the data, to the Standard Model expectation.}
\end{center}
\end{sidewaysfigure}

Distributions of the optimal variables for the couplings
$h_{1}^{\Zo}$ and $h_{4}^{\gamma}$ are shown  in Figure~\ref{fig:4}.
 The regions of
maximal sensitivity to the existence of anomalous couplings correspond to the
largest absolute values of the optimal variables, where discrepancies with
 Standard Model predictions are expected to be larger.

 The 95\%  CL limits on each individual anomalous coupling,
 combining both channels 
and from all data collected at $189\GeV \leq \sqrt{s} \leq 209\GeV$, are obtained
from one-dimensional fits. The results are given in Table \ref{tab:5} and they correspond
to the following intervals:

\begin{center}
\begin{tabular}{lcl}
   $-0.153 < h_{1}^{\Zo}  < 0.141$ & ~~ & $-0.057 < h_{1}^\gamma < 0.057$   \\
   $-0.087 < h_{2}^{\Zo}  < 0.079$ & ~~ & $-0.050 < h_{2}^\gamma < 0.023$   \\
   $-0.220 < h_{3}^{\Zo}  < 0.112$ & ~~ & $-0.059 < h_{3}^\gamma < 0.004$   \\
   $-0.068 < h_{4}^{\Zo}  < 0.148$ & ~~ & $-0.004 < h_{4}^\gamma < 0.042$  \\
\end{tabular}
\end{center}

 To obtain these intervals one parameter is left free at a time, setting the other
 seven anomalous couplings to zero.
 These limits supersede the previous L3 results, obtained
 with a smaller data sample at
 lower centre-of-mass energies\cite{l3}.
 The observed limits agree within 10\% with the expected limits.

  Limits coming from even couplings, $h_{2}^{V}$ and $h_{4}^{V}$
 are more stringent than those coming from odd couplings, $h_{1}^{V}$ and
 $h_{3}^{V}$, as the former correspond to
 operators of dimension eight, while the latter correspond to dimension six
  operators\cite{renard,barroso}, as reflected in equations (1) and (2) in the
  different dependence of the parameters on the mass $m_{\Zo}$ and the
  energy scales. These relations imply that
  the Standard Model is tested more stringently
  in the linear expansion of the effective Lagrangian when
 considering the even couplings.

\begin{figure}
\begin{center}
\begin{tabular}{cc}
\hspace*{-0.5truecm} \includegraphics[width=9.0truecm]
{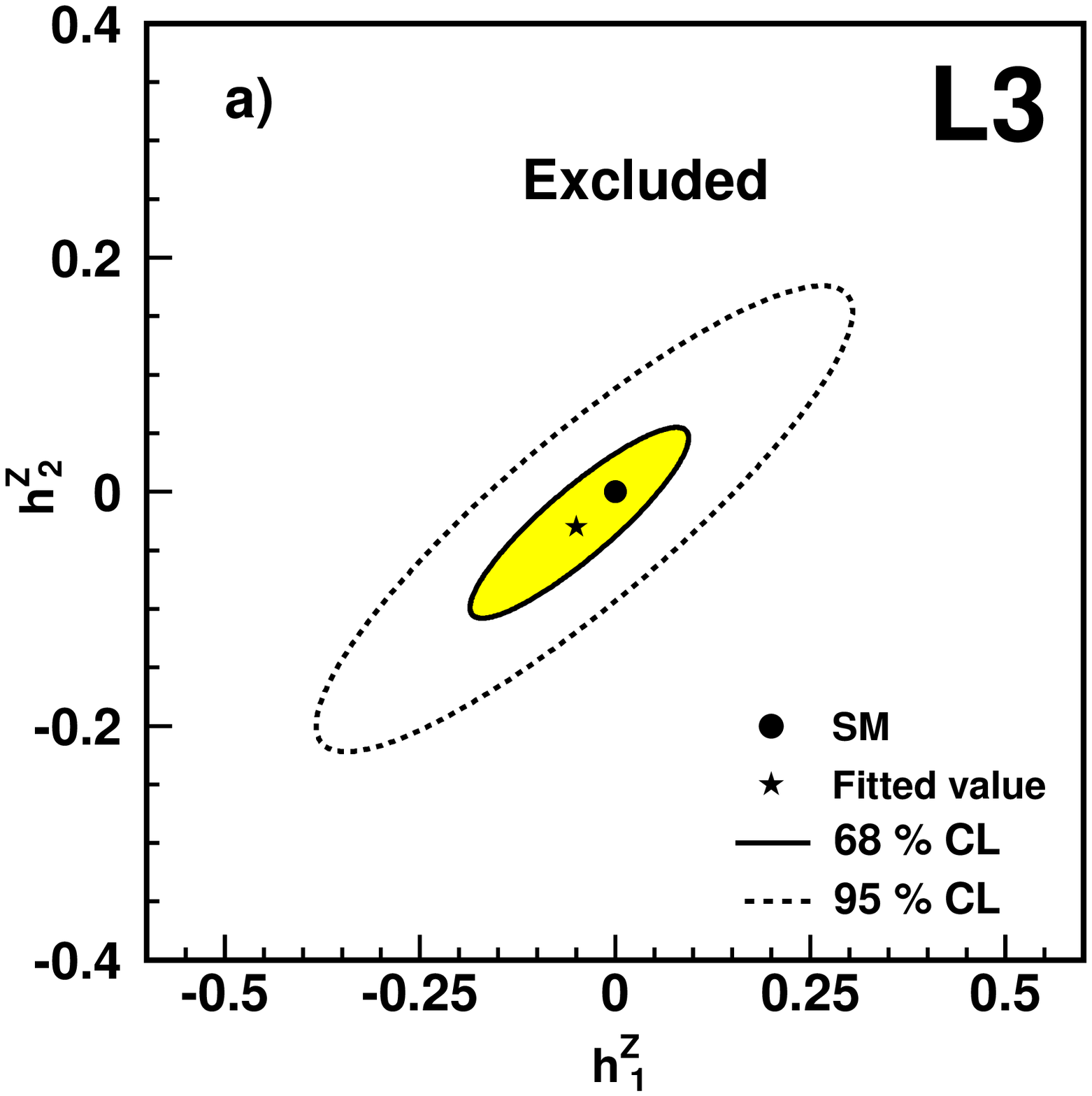} &
\hspace*{-0.5truecm} \includegraphics[width=9.0truecm]
{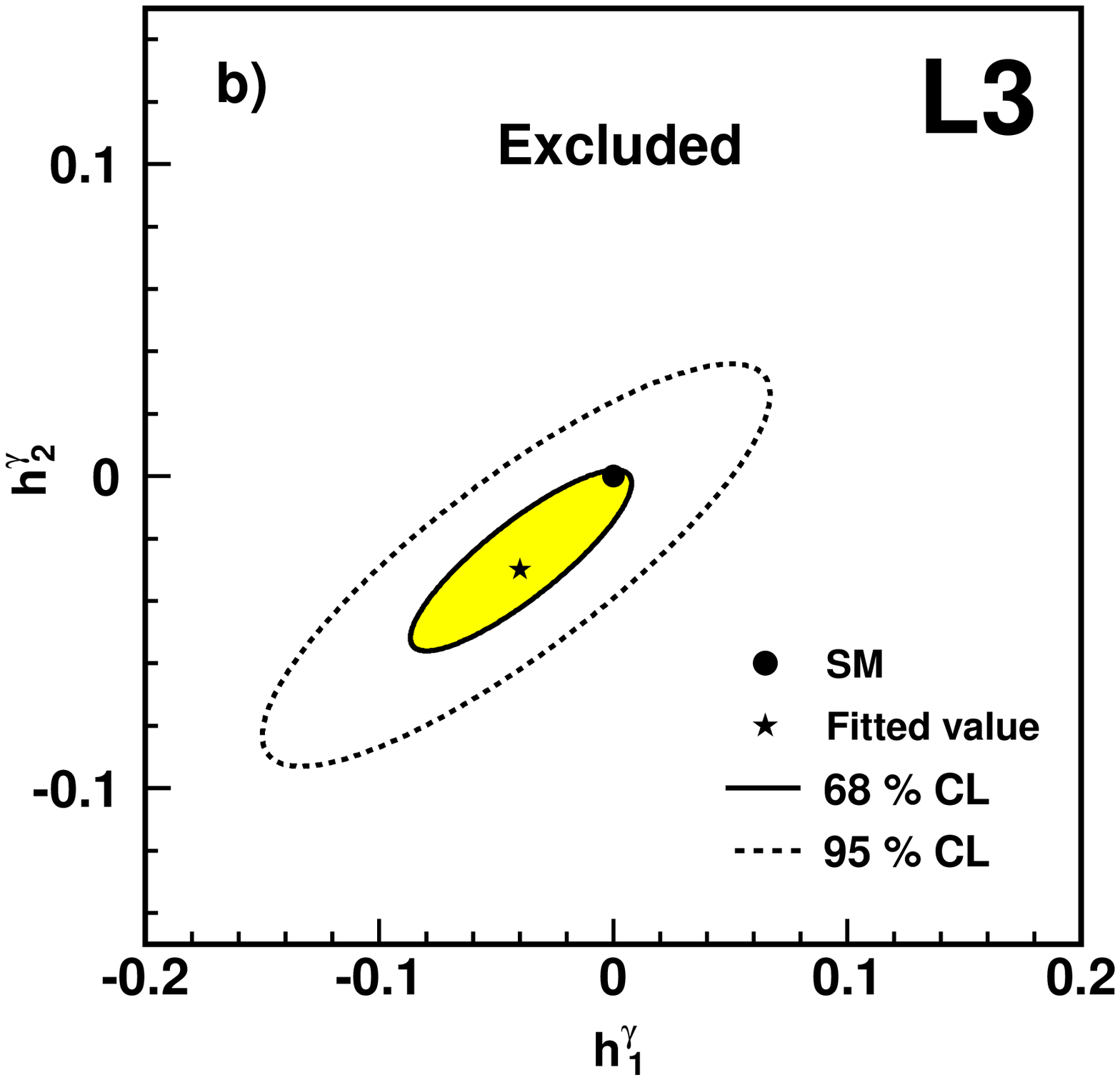} \\
\hspace*{-0.5truecm} \includegraphics[width=9.0truecm]
{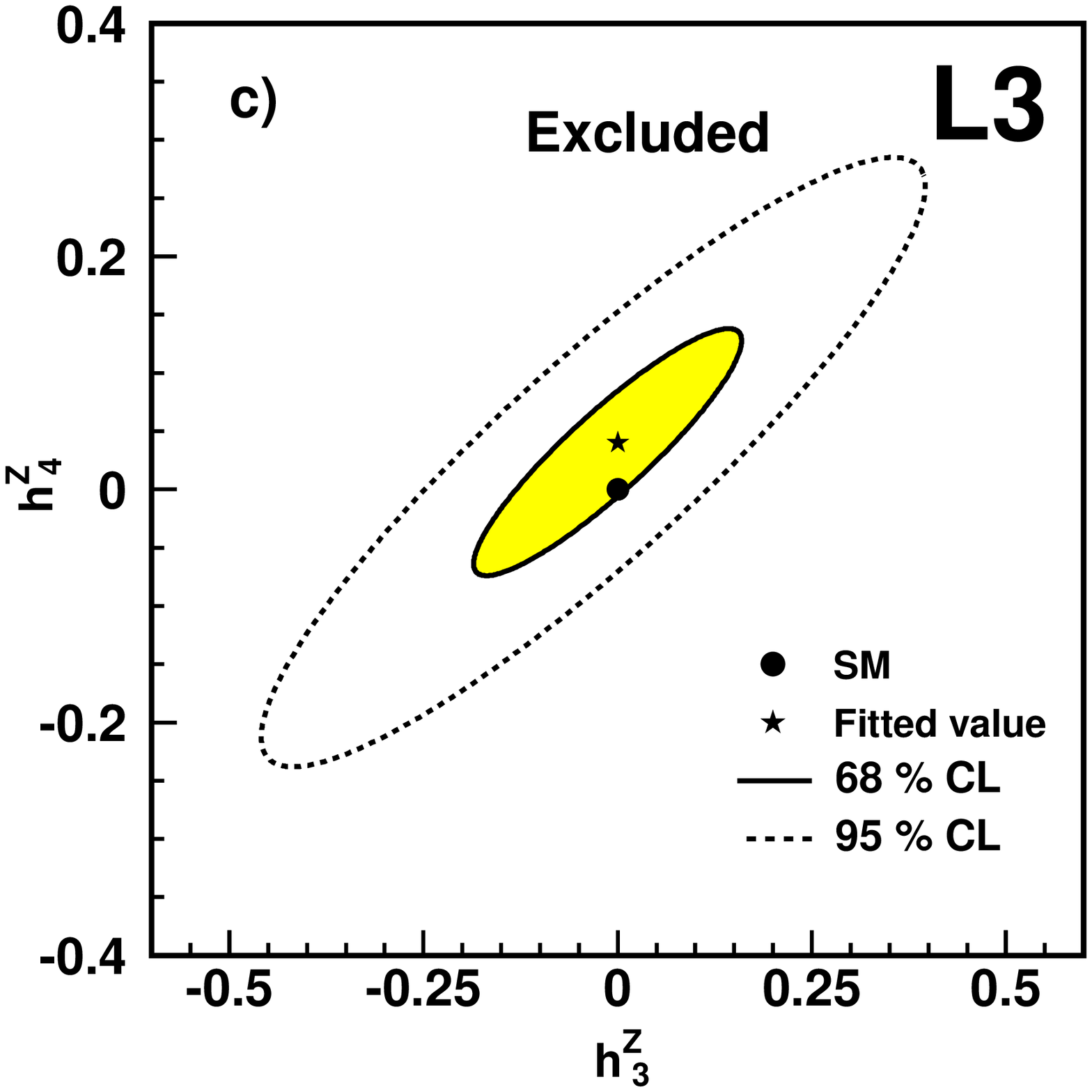} &
\hspace*{-0.5truecm} \includegraphics[width=9.0truecm]
{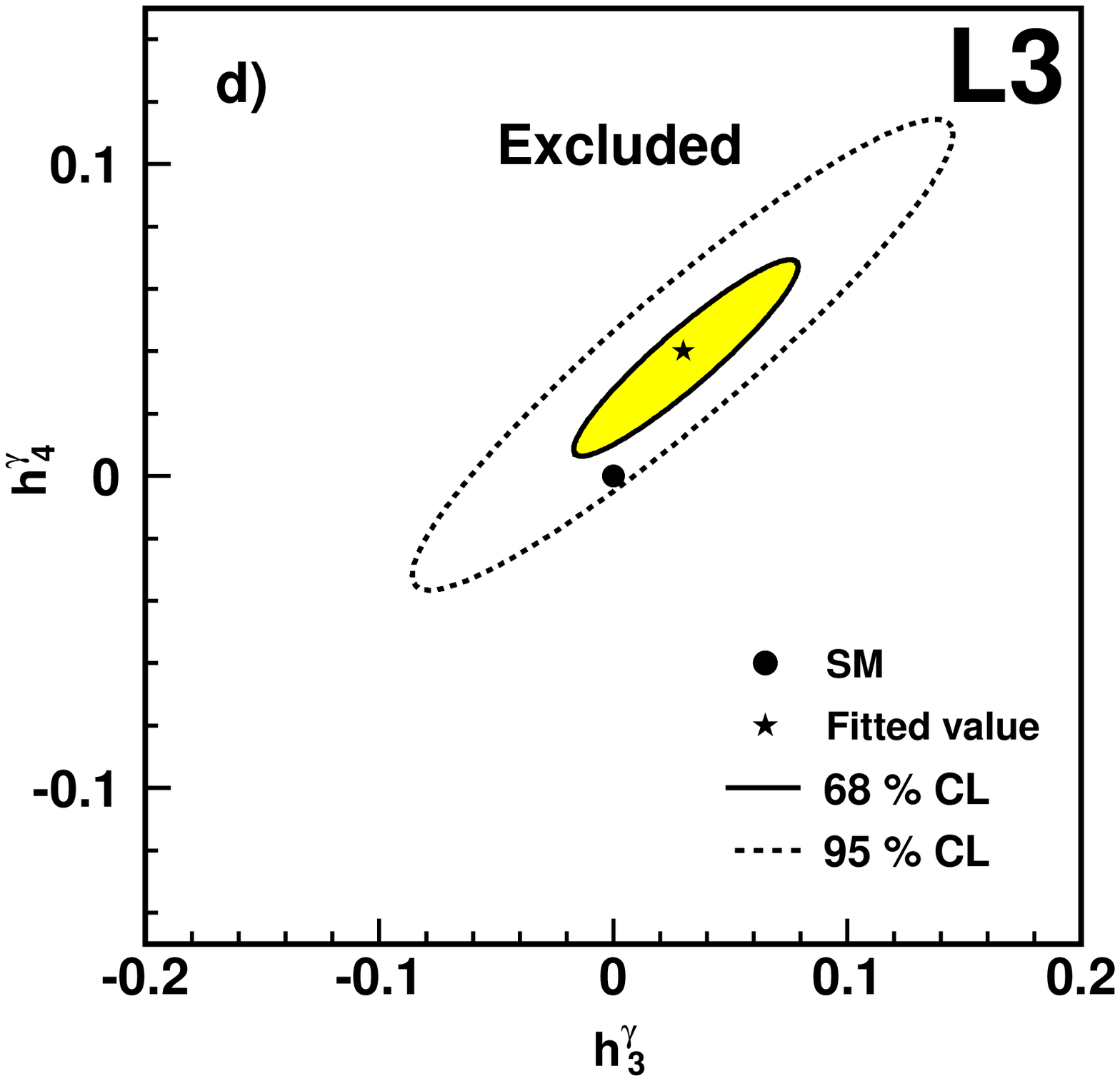} \\
\end{tabular}
\icaption{\label{fig:5}
    Two dimensional limits at 68\% and $95\%$ CL on the pairs of CP-violating
  coupling parameters, a) $h_{2}^{\Zo}$ {\it vs.} $h_{1}^{\Zo}$ and b)
  $h_{2}^{\gamma}$ {\it vs.} $h_{1}^{\gamma}$ and the pairs of CP-conserving coupling
  parameters, c) $h_{4}^{\Zo}$ {\it vs.} $h_{3}^{\Zo}$ and d) $h_{4}^{\gamma}$
  {\it vs.} $h_{3}^{\gamma}$. The Standard Model predictions are indicated
  by the points. The shaded areas correspond to the regions allowed at
  68\% CL while the dashed line shows the 95\% CL exclusion contour.}
\end{center}
\end{figure}

Fits to the two-dimensional distributions of the optimal observables are performed
to determine the pairs of CP-violating ($h_{1}^{V},h_{2}^{V}$)
and CP-conserving couplings ($h_{3}^{V},h_{4}^{V}$),  keeping 
in each case the other couplings fixed at zero. Results 
at $95\%$ CL are shown in Table \ref{tab:6}. A strong correlation between
the two CP-violating or CP-conserving parameters is observed.
Contours for the 68\% and 95\% CL two-dimensional limits on each pair of 
couplings are shown in Figure \ref{fig:5}.
  The main sources of systematic uncertainties, discussed
 in section 4, are included in the limits calculation.
 They contribute 0.02 to one-dimensional limits and 0.03 for two-dimensional 
 limits.

 If the data are interpreted in terms of new physics scales using formulae
(\ref{for:lambda13}) and (\ref{for:lambda24}), lower limits at
 95\% CL on the scale of new physics are obtained as:

\begin{center}
 \begin{tabular}{lcl}
 $\Lambda_{1\Zo} > 0.8 \TeV $ & ~~ & $\Lambda_{1\gamma} > 1.3 \TeV$   \\
 $\Lambda_{2\Zo} > 0.3 \TeV $ & ~~ & $\Lambda_{2\gamma} > 0.4 \TeV$ \\
 $\Lambda_{3\Zo} > 0.8 \TeV $ & ~~ & $\Lambda_{3\gamma} > 2.3 \TeV$ \\
 $\Lambda_{4\Zo} > 0.3 \TeV $ & ~~ & $\Lambda_{4\gamma} > 0.4 \TeV$.  \\
 \end{tabular}
\end{center}

 To determine the confidence levels the probability
distributions are normalized over the physical range of the parameters,
$\Lambda > 0$.


\section{Conclusions}

   The analysis of the process $\epem \ra \Zo\gamma$ in the final states 
 $\ZGQQG$ and $\ZGNNG$ with 620~pb$^{-1}$ of luminosity collected by the L3
 detector at $189 \le \sqrt{s} \le 209 \GeV$ reveals a very good
 agreement between the measured cross sections and the Standard Model prediction.
 Detailed information on the hadronic final-state events are given in form of tables
 to allow constraints of future models.

   These measurements establish upper limits at 95 \% CL on the values of
 anomalous couplings, $h_{i}^{V}$, appearing in the triple neutral-gauge-boson
 vertices, $\Zo\Zo\gamma$ and $\Zo\gamma\gamma$.
  At tree level in the Standard Model these couplings are zero. We observe no
  deviation from this prediction and constrain possible values of the anomalous
  couplings in intervals of widths between 0.05 and 0.33, depending on the coupling
  considered. These limits improve and supersede our previous limits\cite{l3}.

%
\bibliographystyle{l3style}
\bibliography{zg_paper04}
%
%
\newpage
\typeout{   }     
\typeout{Using author list for paper 281 -  }
\typeout{$Modified: Jul 15 2001 by smele $}
\typeout{!!!!  This should only be used with document option a4p!!!!}
\typeout{   }
%
%
%
%
%
%

\newcount\tutecount  \tutecount=0
\def\tutenum#1{\global\advance\tutecount by 1 \xdef#1{\the\tutecount}}
\def\tute#1{$^{#1}$}
\tutenum\aachen            
\tutenum\nikhef            
\tutenum\mich              
\tutenum\lapp              
\tutenum\basel             
\tutenum\lsu               
\tutenum\beijing           
\tutenum\bologna           
\tutenum\tata              
\tutenum\ne                
\tutenum\bucharest         
\tutenum\budapest          
\tutenum\mit               
\tutenum\panjab            
\tutenum\debrecen          
\tutenum\dublin            
\tutenum\florence          
\tutenum\cern              
\tutenum\wl                
\tutenum\geneva            
\tutenum\hamburg           
\tutenum\hefei             
\tutenum\lausanne          
\tutenum\lyon              
\tutenum\madrid            
\tutenum\florida           
\tutenum\milan             
\tutenum\moscow            
\tutenum\naples            
\tutenum\cyprus            
\tutenum\nymegen           
\tutenum\caltech           
\tutenum\perugia           
\tutenum\peters            
\tutenum\cmu               
\tutenum\potenza           
\tutenum\prince            
\tutenum\riverside         
\tutenum\rome              
\tutenum\salerno           
\tutenum\ucsd              
\tutenum\sofia             
\tutenum\korea             
\tutenum\taiwan            
\tutenum\tsinghua          
\tutenum\purdue            
\tutenum\psinst            
\tutenum\zeuthen           
\tutenum\eth               

{
\parskip=0pt
\noindent
{\bf The L3 Collaboration:}
\ifx\selectfont\undefined
 \baselineskip=10.8pt
 \baselineskip\baselinestretch\baselineskip
 \normalbaselineskip\baselineskip
 \ixpt
\else
 \fontsize{9}{10.8pt}\selectfont
\fi
\medskip
\tolerance=10000
\hbadness=5000
\raggedright
\hsize=162truemm\hoffset=0mm
\def\r{\rlap,}
\noindent

P.Achard\r\tute\geneva\ 
O.Adriani\r\tute{\florence}\ 
M.Aguilar-Benitez\r\tute\madrid\ 
J.Alcaraz\r\tute{\madrid}\ 
G.Alemanni\r\tute\lausanne\
J.Allaby\r\tute\cern\
A.Aloisio\r\tute\naples\ 
M.G.Alviggi\r\tute\naples\
H.Anderhub\r\tute\eth\ 
V.P.Andreev\r\tute{\lsu,\peters}\
% ...
\newpage
\begin{list}{A}{\itemsep=0pt plus 0pt minus 0pt\parsep=0pt plus 0pt minus 0pt
                \topsep=0pt plus 0pt minus 0pt}
\item[\aachen]
 III. Physikalisches Institut, RWTH, D-52056 Aachen, Germany$^{\S}$
\item[\nikhef] National Institute for High Energy Physics, NIKHEF, 
     and University of Amsterdam, NL-1009 DB Amsterdam, The Netherlands
\item[\mich] University of Michigan, Ann Arbor, MI 48109, USA
\item[\lapp] Laboratoire d'Annecy-le-Vieux de Physique des Particules, 
     LAPP,IN2P3-CNRS, BP 110, F-74941 Annecy-le-Vieux CEDEX, France
\item[\basel] Institute of Physics, University of Basel, CH-4056 Basel,
     Switzerland
\item[\lsu] Louisiana State University, Baton Rouge, LA 70803, USA
\item[\beijing] Institute of High Energy Physics, IHEP, 
  100039 Beijing, China$^{\triangle}$ 
\item[\bologna] University of Bologna and INFN-Sezione di Bologna, 
     I-40126 Bologna, Italy
\item[\tata] Tata Institute of Fundamental Research, Mumbai (Bombay) 400 005, India
\item[\ne] Northeastern University, Boston, MA 02115, USA
\item[\bucharest] Institute of Atomic Physics and University of Bucharest,
     R-76900 Bucharest, Romania
\item[\budapest] Central Research Institute for Physics of the 
     Hungarian Academy of Sciences, H-1525 Budapest 114, Hungary$^{\ddag}$
\item[\mit] Massachusetts Institute of Technology, Cambridge, MA 02139, USA
\item[\panjab] Panjab University, Chandigarh 160 014, India
\item[\debrecen] KLTE-ATOMKI, H-4010 Debrecen, Hungary$^\P$
\item[\dublin] Department of Experimental Physics,
  University College Dublin, Belfield, Dublin 4, Ireland
\item[\florence] INFN Sezione di Firenze and University of Florence, 
     I-50125 Florence, Italy
\item[\cern] European Laboratory for Particle Physics, CERN, 
     CH-1211 Geneva 23, Switzerland
\item[\wl] World Laboratory, FBLJA  Project, CH-1211 Geneva 23, Switzerland
\item[\geneva] University of Geneva, CH-1211 Geneva 4, Switzerland
\item[\hamburg] University of Hamburg, D-22761 Hamburg, Germany
\item[\hefei] Chinese University of Science and Technology, USTC,
      Hefei, Anhui 230 029, China$^{\triangle}$
\item[\lausanne] University of Lausanne, CH-1015 Lausanne, Switzerland
\item[\lyon] Institut de Physique Nucl\'eaire de Lyon, 
     IN2P3-CNRS,Universit\'e Claude Bernard, 
     F-69622 Villeurbanne, France
\item[\madrid] Centro de Investigaciones Energ{\'e}ticas, 
     Medioambientales y Tecnol\'ogicas, CIEMAT, E-28040 Madrid,
     Spain${\flat}$ 
\item[\florida] Florida Institute of Technology, Melbourne, FL 32901, USA
\item[\milan] INFN-Sezione di Milano, I-20133 Milan, Italy
\item[\moscow] Institute of Theoretical and Experimental Physics, ITEP, 
     Moscow, Russia
\item[\naples] INFN-Sezione di Napoli and University of Naples, 
     I-80125 Naples, Italy
\item[\cyprus] Department of Physics, University of Cyprus,
     Nicosia, Cyprus
\item[\nymegen] University of Nijmegen and NIKHEF, 
     NL-6525 ED Nijmegen, The Netherlands
\item[\caltech] California Institute of Technology, Pasadena, CA 91125, USA
\item[\perugia] INFN-Sezione di Perugia and Universit\`a Degli 
     Studi di Perugia, I-06100 Perugia, Italy   
\item[\peters] Nuclear Physics Institute, St. Petersburg, Russia
\item[\cmu] Carnegie Mellon University, Pittsburgh, PA 15213, USA
\item[\potenza] INFN-Sezione di Napoli and University of Potenza, 
     I-85100 Potenza, Italy
\item[\prince] Princeton University, Princeton, NJ 08544, USA
\item[\riverside] University of Californa, Riverside, CA 92521, USA
\item[\rome] INFN-Sezione di Roma and University of Rome, ``La Sapienza",
     I-00185 Rome, Italy
\item[\salerno] University and INFN, Salerno, I-84100 Salerno, Italy
\item[\ucsd] University of California, San Diego, CA 92093, USA
\item[\sofia] Bulgarian Academy of Sciences, Central Lab.~of 
     Mechatronics and Instrumentation, BU-1113 Sofia, Bulgaria
\item[\korea]  The Center for High Energy Physics, 
     Kyungpook National University, 702-701 Taegu, Republic of Korea
\item[\taiwan] National Central University, Chung-Li, Taiwan, China
\item[\tsinghua] Department of Physics, National Tsing Hua University,
      Taiwan, China
\item[\purdue] Purdue University, West Lafayette, IN 47907, USA
\item[\psinst] Paul Scherrer Institut, PSI, CH-5232 Villigen, Switzerland
\item[\zeuthen] DESY, D-15738 Zeuthen, Germany
\item[\eth] Eidgen\"ossische Technische Hochschule, ETH Z\"urich,
     CH-8093 Z\"urich, Switzerland
\item[\S]  Supported by the German Bundesministerium 
        f\"ur Bildung, Wissenschaft, Forschung und Technologie.
\item[\ddag] Supported by the Hungarian OTKA fund under contract
numbers T019181, F023259 and T037350.
\item[\P] Also supported by the Hungarian OTKA fund under contract
  number T026178.
\item[$\flat$] Supported also by the Comisi\'on Interministerial de Ciencia y 
        Tecnolog{\'\i}a.
\item[$\sharp$] Also supported by CONICET and Universidad Nacional de La Plata,
        CC 67, 1900 La Plata, Argentina.
\item[$\triangle$] Supported by the National Natural Science
  Foundation of China.
\end{list}
}
\vfill


%
%
\newpage
\begin{table}[t]
\centering
\begin{tabular}{|c|c|c|c|c|c|c|}
\hline
   $\sqrt{s}~~({\rm GeV})$ & ${\cal L}$ ($\pb$) & $\epsilon$ (\%) & Back. (\%) &
  Events  & $\sigma$ (pb)           & $\sigma_{SM}$ (pb)\\
\hline

 188.6 & 172.1 & 28.9 $\pm$ 0.1 & 0.9 & $899$ & 18.1 $\pm$ 0.6 $\pm$ 0.2 & 18.7 $\pm$ 0.2 \\
 191.6 & $\;\; 17.9$ & 28.2 $\pm$ 0.4 & 0.8 & $101$ & 20.0 $\pm$ 2.1 $\pm$ 0.2  & 18.8 $\pm$ 0.2 \\
 195.5 & $\;\; 74.9$ & 27.1 $\pm$ 0.2 & 0.9 & $351$ & 17.3 $\pm$ 1.1 $\pm$ 0.2 & 17.9 $\pm$ 0.2 \\
 199.5 & $\;\; 67.4$ & 27.3 $\pm$ 0.2 & 1.0 & $333$ & 18.1 $\pm$ 1.0 $\pm$ 0.2 & 16.9 $\pm$ 0.2 \\
 201.7 & $\;\; 36.5$ & 27.4 $\pm$ 0.3 & 1.0 & $163$ & 16.3 $\pm$ 1.3 $\pm$ 0.2 & 16.3 $\pm$ 0.2 \\
 $202.5 - 205.5$ & $\;\; 78.7$  & 25.0 $\pm$ 0.1 & 4.8 & $325$ & 16.4  $\pm$ 1.0 $\pm$ 0.2 & 16.3 $\pm$ 0.2 \\
 $205.5 - 207.2$ & 124.1  & 25.2 $\pm$ 0.1 & 4.7 & $494$ & 15.8  $\pm$ 0.7 $\pm$ 0.2 & 16.1 $\pm$ 0.2 \\
 $207.2 - 209.2$ & $\;\;\;\; 8.2$ & 24.8 $\pm$ 0.1 & 4.7 & $\;\; 33$ & 16.1 $\pm$ 3.0 $\pm$ 0.2 & 15.6 $\pm$ 0.2 \\
\hline
\end{tabular}
\icaption{ Integrated luminosities, ${\cal L}$, and
results of the $\EEQQG$ selection:
           selection efficiencies, $\epsilon$,
     background level,
    number of selected events (background subtracted) and measured cross sections
    with statistical and systematic uncertainties. The  uncertainty on $\epsilon$ comes from
    Monte Carlo statistics. The corresponding Standard Model cross sections, 
    $\sigma_{SM}$, are listed in the last column.
    They are derived from the KK2f Monte Carlo generator {\protect \cite{kk2f}}.
    Their uncertainty includes a 1\% theory uncertainty and finite Monte Carlo
          statistics.
\label{tab:1}}
\end{table}

%
%
\begin{table}[t]
\centering
\begin{tabular}{|c|c|c|c|c|c|}
\hline

  $\sqrt{s}~~({\rm GeV})$ & ${\cal L}$ ($\pb$) & $\epsilon$ (\%) & 
  Events  & $\sigma$ (pb) & $\sigma_{\rm SM}$ (pb)\\
\hline
 188.6 & $175.6$ & 32.3 $\pm$ 0.4 &  $288$ &  5.1 $\pm$ 0.3 $\pm$ 0.1 &
      4.99 $\pm$ 0.05 \\
 191.6 & $\;\; 16.9$ & 31.4 $\pm$ 0.5 & $\;\; 25$ &  4.7 $\pm$ 0.8 $\pm$ 0.1 &
      4.85 $\pm$ 0.05 \\
 195.5 & $\;\; 80.9$ & 31.5 $\pm$ 0.4 &  $107$ &  4.2 $\pm$ 0.4 $\pm$ 0.1 &
      4.57 $\pm$ 0.05 \\
 199.5 & $\;\; 79.5$ & 28.0 $\pm$ 0.4 & $\;\; 98$ & 4.4 $\pm$ 0.4 $\pm$ 0.1 &
      4.42 $\pm$ 0.05 \\
 201.7 & $\;\; 36.1$ & 30.8 $\pm$ 0.4 &  $\;\; 50$ &  4.5 $\pm$ 0.6 $\pm$ 0.1 &
      4.34 $\pm$ 0.04 \\
 $202.5 - 205.5$ & $\;\; 74.3$  & 28.9 $\pm$ 0.5 & $\;\; 88$ &  4.1  $\pm$ 0.4 $\pm$ 0.1 &
      4.17 $\pm$ 0.04 \\
 $205.5 - 209.2$ & $129.6$  & 29.4 $\pm$ 0.5 &  $160$ &  4.2  $\pm$ 0.3 $\pm$ 0.1 &
      4.09 $\pm$ 0.04 \\
\hline
\end{tabular}
\caption{Integrated luminosities, ${\cal L}$, and results of the 
$\EENNG$ selection: selection efficiencies, $\epsilon$, number of selected events and
 measured cross sections with statistical and systematic uncertainties. 
  The uncertainty on $\epsilon$ comes from Monte Carlo statistics. The
 corresponding Standard Model cross sections, $\sigma_{\rm SM}$,
 are listed in the last column. They are derived from the KKMC Monte Carlo 
 generator {\protect \cite{kk2f}}. Their uncertainty includes a 1\% theory
          uncertainty and a contribution from finite Monte Carlo statistics.
\label{tab:2}}
\end{table}
%
%
\begin{sidewaystable}
 \begin{center}
\begin{tabular}{|c|ccccc|}
 \cline{2-6} 
 \multicolumn{1}{c}{  } & \multicolumn{5}{|c|}{$m_{rec}$ [$\!\!\GeV$] }  \\
 \cline{1-1}
 \rule{0pt}{12pt}$|\cos\theta_{\gamma}|$ & $80 - 88$ & $88 - 92$ & $92 - 96$ & $96 - 104$ &
  $104 - 110$ \\
\hline
 $0.00 - 0.20$ & $14/20.4/0.0/78$ & $\;\;64/\;\;69.8/\;\;0.0/35$ & 
 $\;\;79/\;\;62.3/\;\;0.0/35$  & $\;\;58/\;\;55.8/\;\;0.5/57$ & $43/24.0/0.3/67$ \\
 $0.20 - 0.40$ & $19/27.1/0.0/85$ & $\;\;91/\;\;86.4/\;\;0.0/37$ &
 $\;\;92/\;\;75.4/\;\;0.9/38$  & $\;\;62/\;\;68.2/\;\;0.3/62$ & $31/26.9/0.7/64$ \\
 $0.40 - 0.60$ & $26/25.7/0.0/62$ & $116/110.1/\;\;0.4/37$ & $121/107.9/\;\;0.5/41$ &
 $\;\;79/\;\;83.8/\;\;0.6/64$ & $35/32.9/0.6/64$ \\
 $0.60 - 0.80$ & $29/19.3/0.2/28$ & $\;\;88/\;\;84.1/\;\;0.5/17$ & 
 $\;\;75/100.6/\;\;0.1/24$  & $\;\;82/\;\;92.3/\;\;1.3/43$ & $37/40.7/0.5/48$ \\
 $0.80 - 0.90$ & $35/39.7/0.6/61$ & $145/167.4/\;\;0.4/35$ & $127/159.6/\;\;1.9/40$
  & $161/136.1/\;\;4.3/63$ & $68/55.7/1.9/70$ \\
 $0.90 - 0.99$ & $95/55.9/4.6/18$ & $302/276.8/11.5/12$ & $271/281.9/11.8/15$ &
 $232/225.8/12.3/24$ & $98/87.1/4.4/24$ \\
\hline
\end{tabular}
 \icaption{Numbers of events selected in the $\EEQQG$ channel,
  Standard Model expectations, background level and selection
        efficiencies (in \%) as a function of the recoil mass, $m_{rec}$ to the
       photon candidate and of the absolute value of the cosine of the
        photon polar angle, $|\cos\theta_{\gamma}|$.
 \label{tab:3}}
\end{center}
\end{sidewaystable}
%
%
\newpage 
\begin{table}[t]
\centering
\begin{tabular}{|c|c|c|}
\hline
     & \multicolumn{2}{|c|}{Uncertainty (\%)}  \\
\cline{2-3}
    Source     &  $\EEQQG$   & $\EENNG$    \\
\hline
 Selection Process   & 0.8  & 1.5  \\
 Photon energy scale & 0.4  &  0.6   \\
 MC statistics       & $ \!\!\!\!\!\!\le$ 0.4 &  0.4   \\
 Luminosity          & 0.2  & 0.2  \\
 Trigger efficiency  &  --  & 0.3 \\
 Background level    & 0.3 & -- \\
\hline
 Total &  1.1 & 1.7 \\
\hline
\end{tabular}
\icaption{ Sources of systematic uncertainty in the 
 \mbox{$\EEQQG$} and \mbox{$\EENNG$} cross sections.
\label{tab:4}}
\end{table}
%
%
\begin{table}[t]
\centering
\begin{tabular}{|c|c|c|c|}
\hline
   ~~       & Fitted & Negative & Positive  \\
  Parameter & value  & error ($95 \% \, {\rm CL}$) & error ($95 \% \, {\rm CL}$) \\
\hline

 \rule{0pt}{12pt} $h_{1}^{\Zo}$ &  $ -0.007$ & $ 0.146$ & $0.148$  \\
 \rule{0pt}{12pt} $h_{2}^{\Zo}$ &  $ -0.006$ & $ 0.080$ & $0.085$  \\
 \rule{0pt}{12pt} $h_{3}^{\Zo}$ &  $ -0.036$ & $ 0.184$ & $0.148$  \\
 \rule{0pt}{12pt} $h_{4}^{\Zo}$ &  $ \phantom{-}0.038$ & $ 0.106$ & $0.110$  \\
 \rule{0pt}{12pt} $h_{1}^{\gamma}$ &  $-0.001$ & $ 0.056$ & $0.058$  \\
 \rule{0pt}{12pt} $h_{2}^{\gamma}$ &  $-0.014$ & $ 0.035 $ & $0.037$   \\
 \rule{0pt}{12pt} $h_{3}^{\gamma}$ &  $-0.026$ & $ 0.033$ & $0.031$  \\
 \rule{0pt}{12pt} $h_{4}^{\gamma}$ &  $\phantom{-}0.020$ & $ 0.024$ & $0.022$  \\
\hline
\end{tabular}
\icaption{ Fitted values and errors at $95$ \% CL on the individual anomalous couplings
 from one-dimensional fits. In each fit the other seven parameters are set to zero.
\label{tab:5}}
\end{table}
%
%
%
%
\begin{table}[t]
\centering
\begin{tabular}{|c|c|c|c|c|}
\hline
   ~~       & Fitted & Negative & Positive & Correlation \\
  Parameter & value  & limits   & limits   & coeficient  \\
\hline

 \rule{0pt}{12pt} $h_{1}^{\Zo}$ &  $ -0.05$ & $-0.38$ & $0.30$ & 0.89 \\
 \rule{0pt}{12pt} $h_{2}^{\Zo}$ &  $ -0.03$ & $-0.22$ & $0.18$ & ~~ \\
\hline
 \rule{0pt}{12pt} $h_{3}^{\Zo}$ &  $ -0.00$ & $-0.46$ & $0.40$ & 0.90 \\
 \rule{0pt}{12pt} $h_{4}^{\Zo}$ &  $ \phantom{-}0.04$ & $-0.24$ & $0.28$ & ~~ \\
\hline
 \rule{0pt}{12pt} $h_{1}^{\gamma}$ &  $-0.04$ & $-0.15$ & $0.07$ & 0.85 \\
 \rule{0pt}{12pt} $h_{2}^{\gamma}$ &  $-0.03$ & $-0.09 $ & $0.04$ & ~~ \\
\hline
 \rule{0pt}{12pt} $h_{3}^{\gamma}$ &  $\phantom{-}0.03$ & $-0.09$ & $0.14$ & 0.93 \\
 \rule{0pt}{12pt} $h_{4}^{\gamma}$ &  $\phantom{-}0.04$ & $-0.04$ & $0.11$ & ~~ \\
\hline
\end{tabular}
\icaption{ Limits at $95$ \% CL on pairs of anomalous couplings from  
two-dimensional fits. In each fit the other six parameters are set to zero.
\label{tab:6}}
\end{table}
%
%
%
%
%
%
\newpage
%
%
 \newpage
\newpage
%
%
\newpage
%
\newpage
\end{document}